\documentclass[11pt,preprintnumbers,aps,amssymb,nofootinbib,amsmath,superscriptaddress,notitlepage]{revtex4-1}
\usepackage{epsfig,epsf}
\usepackage{bm} 
\usepackage{color} 
\usepackage{slashed}
\usepackage{relsize}	
\usepackage{soul} 
\usepackage{hyperref}
\newcommand{\beq}{\begin{equation}}
\newcommand{\beql}[1]{\begin{equation}\label{#1}}
\newcommand{\eeq}{\end{equation}}
\def\bal#1\gal{\begin{align}#1\end{align}}
\newcommand{\ball}[1]{\bal\label{#1}}
\newcommand{\eq}[1]{(\ref{#1})}
\newcommand{\fig}[1]{Fig.~\ref{#1}}

\renewcommand{\b}[1]{{\bm #1}} 
\newcommand{\unit}[1]{\hat {{\bm #1}}} 

\begin{document}

%
\title{Continuous evolution of electromagnetic field in heavy-ion collisions}

\author{Evan Stewart}
\author{Kirill Tuchin}

\affiliation{Department of Physics and Astronomy, Iowa State University, Ames, Iowa, 50011, USA}

\date{\today}


\begin{abstract}

In heavy-ion collisions the electromagnetic field exists before the hot nuclear matter emergence. Requiring the field continuity we compute it in the central rapidity region by taking into account the electromagnetic response of the Quark Gluon Plasma. We show that the electromagnetic field is nearly time-independent from about 1~fm/c after the collision until the freezeout.

\end{abstract}

\maketitle


Relativistic heavy-ion collisions produce arguably the most intense electromagnetic field in nature \cite{Kharzeev:2007jp,Skokov:2009qp,Voronyuk:2011jd,Ou:2011fm,Bzdak:2011yy,Bloczynski:2012en,Deng:2012pc}. This field is of great interest as it probes the intensity frontier of QED and induces novel effects in Quark  Gluon Plasma (QGP) \cite{Kharzeev:2013jha}. It is produced   mostly by the valence electric charges of the original heavy-ions and by the currents induced in QGP. Its dynamics approximately decouple from that of the hot nuclear medium, owing to the relative weakness of the electromagnetic interactions as compared to the strong ones. Thus one can study independently the effect of the  background electromagnetic field on QGP \cite{Roy:2017yvg,Pu:2016ayh,Roy:2015kma,Pu:2016bxy,Roy:2015coa,Roy:2017yvg,Inghirami:2016iru,Mohapatra:2011ku,Das:2017qfi,Greif:2017irh,Tuchin:2011jw} and the effect of the electromagnetic response of QGP, its spacetime distribution and motion on the electromagnetic field
\cite{Tuchin:2010vs,Tuchin:2013apa,Tuchin:2013ie,Zakharov:2014dia,Tuchin:2015oka,Li:2016tel,Gursoy:2014aka,Gursoy:2018yai,Stewart:2017zsu}.  The latter is the subject of this work. 

The electromagnetic field before the collision at $t=0$ (in the laboratory frame) is sourced by the boosted Coulomb fields of the heavy-ions. The collision sends the color excitations of the heavy-ion wave functions onto the mass-shell by the time $t_0=1/Q_s$, where $Q_s$ is the saturation momentum. These excitations are produced mostly in the transverse direction relative to the collision axis and make up the bulk of the QGP that occupies the interaction volume. However, the valence quarks experience very little deflection from the incident straight-line trajectories  \cite{Kharzeev:1996sq,Itakura:2003jp}. They continue to induce the electromagnetic field at $t>t_0$. Since the interaction region of the heavy-ions is now occupied by the nuclear medium its electromagnetic response induces another contribution to the electromagnetic field which clearly  depends on the properties of QGP. The latter behaves as a near perfect fluid as early as 0.5~fm/$c$ (and in some models even earlier) and its response can be described by electrical conductivity $\sigma$. Under the decoupling hypothesis, the total electromagnetic field in the fluid can be computed by solving the Maxwell equations, which is the standard approach. However, it ignores the fact that at $t\sim t_0$ the electromagnetic field in the nuclear medium must match to the one that existed in the interaction volume before the medium emergence. In \cite{Tuchin:2015oka} a model was proposed that ascribes an effective electrical conductivity $\sigma(t)$ to the nuclear medium at the earliest times so that the solution to the Maxwell equations in the fluid QGP can be extended all the way to $t=t_0$. This is the approach adopted in this paper as well.\footnote{A promising magneto-hydrodynamics approach to the early stages was proposed recently in \cite{Yan:2021zjc}.} Our goal is to compute the electromagnetic field in QGP after a heavy-ion collision by solving the Maxwell equations with the  appropriate initial conditions.

\medskip

Consider an incident proton traveling in vacuum along the $z$-axis, at transverse distance $\b b$ from it,  until $t=t_0$. Let the corresponding field potential be $\vec A_1(\b r, t)$. We use the arrow on top of a symbol to denote a 4-vector, and the bold-face to denote a 3-vector. At $t>t_0$ the proton travels through QGP whose electromagnetic response we describe only by one transport coefficient: the electrical conductivity $\sigma$. As nuclear medium emerges and expands, $\sigma$  changes with time, but for the time being this evolution will be neglected. The electromagnetic potential at this later time will be denoted as $\vec A_2(\b r, t)$. We have
\ball{a1}
 \vec A(\b r, t) =  \vec A_1(\b r, t)\theta(t_0-t)+ \vec A_2(\b r,t)\theta(t-t_0)\,.
\gal
At early times $t<t_0$ the electromagnetic potential in the covariant gauge  $\vec\partial\cdot \vec A_1=0$
obeys the wave equation  $ \partial^2 \vec A_1 =\vec j$, where $\vec j$ is the proton current.  Since we neglect proton deflection in the course of the collision, this current does not depend on $t_0$ and consequently bears no subscript 1 or 2. Solution of the wave equation describing the potential of a proton coming in from infinity can be written in terms of the retarded Green's function
\ball{a3}
G_1(\b r, t|\b r',t')= \frac{1}{4\pi}\frac{\delta(t-t'-|\b r-\b r'|)}{|\b r-\b r'|}\,,
\gal
as
\ball{a5}
\vec A_1(\b r, t) = \int_{-\infty}^{t+}dt'\int d^3r' \vec j(\b r',t')G_1(\b r, t|\b r',t')\,.
\gal 
Evaluated at $t=t_0$ Eq.~\eq{a5} and its time derivative yield the initial conditions $ \vec {\mathcal{A}}(\b r)$ and $\vec {\mathcal{V}}(\b r)$.

Electromagnetic potential at $t>t_0$ is governed by the equations 
\ball{a7}
\partial^2 \vec A_2+ \sigma\partial_t \vec A_2 =\vec j \,,
\gal
provided that the potentials are constrained by the gauge condition
\ball{a9}
\vec\partial \cdot \vec A_2+ \sigma \varphi_2 = \partial_t\varphi_2 + \b\nabla \cdot \b A_2+\sigma \varphi_2=0\,.
\gal 
Eqs.~\eq{a7} and \eq{a9} are not boost-invariant, as they hold only in the center-of-mass, or the ``lab", frame where the electrical conductivity $\sigma$ is defined.  $\vec A_2 (\b r, t_0)$ must satisfy the initial conditions 
\bal
\vec A_2 (\b r, t_0)&= \vec A_1(\b r, t_0)\equiv \vec {\mathcal{A}}(\b r)\,, \label{a11}\\
\partial_t \vec A_2 (\b r, t)\big|_{t=t_0}&= \partial_t \vec A_1 (\b r, t)\big|_{t=t_0}\equiv \vec {\mathcal{V}}(\b r)\,. \label{a13}
\gal
The desired solution can be expressed in terms of the Green's function $G_2(\b r, t| \b r', t')$ that satisfies the equation 
\begin{subequations}\label{a15}
\bal
&(\partial^2 + \sigma\partial_t) G_2(\b r, t| \b r', t') =\delta(\b r-\b r')\delta(t-t')\,
\gal
with the initial condition 
\bal
G_2(\b r, t| \b r', t') =0\,,\quad \text{if}\quad  t<t'\,.
\gal
\end{subequations}
Solution to the initial value problem  \eq{a15} reads \cite{MF}
\begin{subequations}\label{a17}
\bal
G_2(\b r, t| \b r', t')&= \frac{1}{4\pi}e^{-\frac{1}{2}\sigma(t- t')}\frac{\delta(t-t'-R)}{R}\theta(t-t')\label{a19}\\
&+\frac{1}{4\pi}e^{-\frac{1}{2}\sigma(t- t')}\frac{\sigma/2}{\sqrt{(t-t')^2-R^2}}I_1\left(\frac{\sigma}{2}\sqrt{(t-t')^2-R^2}\right)\theta (t-t'-R)\theta(t-t')\,,\label{a21}
\gal
\end{subequations}
where $ R= |\b r-\b r'|$. The first term \eq{a19} represents  the original pulse, whereas the second term \eq{a21}  the wake created by the currents induced in the plasma. The exponential factor $\exp[-\sigma(t-t')/2]$ indicates the decrease of the field strength due to the work done by the field on the electric currents in the plasma. $I_1$ is a modified Bessel function. It has been argued in \cite{Stewart:2017zsu} that at $t\gtrsim t_0$, i.e.\ soon after QGP emerges,  the pulse term is dominant, while the wake term dominates when $t\gg t_0$. The transitions from the pulse dominated field to the wake dominated field occurs around time $4/\sigma\sim 10^2$~fm for the QGP produced in relativistic heavy-ion collisions. This is an order of magnitude longer that the plasma lifetime. Thus, we are allowed to neglect the wake contribution. Essentially, this means  that the QGP is a poor conductor of electric current. 

Using the Green's function \eq{a17} we can write the solution to the initial value problem \eq{a11},\eq{a13} as
\begin{subequations}\label{a23}
\bal
\vec A_2(\b r, t)=& \int_{t_0}^{t+}dt' \int d^3r'  \vec j (\b r',t')G_2(\b r, t|\b r',t')\label{a25}\\
&+ \int d^3r' \left[ \sigma \vec{\mathcal{A}}(\b r')+\vec{\mathcal{V}}(\b r')\right]
G_2(\b r, t|\b r',t')\big|_{t'=t_0}\label{a26}\\
&-\int d^3r' \vec{ \mathcal{A}}(\b r')\partial_{ t'}G_2(\b r, t|\b r',t')\big|_{t'=t_0}\,.\label{a26x}
\gal
\end{subequations}
Upon substituting \eq{a19} for the Green's function, the contribution of \eq{a23} reads
\begin{subequations}\label{a35}
\ball{e1} 
\vec A_2(\b r, t)&=\vec A_2^\text{val}(\b r, t)+\vec A_2^\text{init}(\b r, t)\,,
\gal
where the part of the total electromagnetic field due to the current of the valence charges at $t>t_0$ is 
\ball{a37}
\vec A_2^\text{val}(\b r, t)&=\frac{1}{4\pi} \int_{t_0}^{t+} \int \frac{\delta(t-t'-R)}{R} e^{-\frac{1}{2}\sigma (t-t')} \theta(t-t')\vec j(\b r',t')d^3r' dt'\,,
\gal
and the part of the total electromagnetic field due to the initial fields at $t=t_0$ is
\bal
\vec A_2^\text{init}(\b r, t)&=\frac{1}{4\pi}\bigg\{ (t-t_0)e^{-\frac{1}{2}\sigma(t- t_0)}\oint_{S_{\b r}^{t-t_0}}  [\sigma \vec{ \mathcal{A}}(\b r)+\vec{\mathcal{V}}(\b r)]d\Omega\label{a38}\\
&+\partial_t\left[(t-t_0)e^{-\frac{1}{2}\sigma(t-t_0)}\oint_{S_{\b r}^{t-t_0}}  \vec{\mathcal{A}}(\b r)d\Omega\right]\bigg\}\,,
\label{a39}
\gal
\end{subequations}
$S_{\b r}^{t-t_0}$ is a sphere of radius $t-t_0$ with the center at $\b r$ and $d\Omega$ is the solid angle element on it. Thus, for example, employing the spherical coordinates  one can write
\ball{a40}
&\oint_{S_{\b r}^{\Delta t}}  \vec{\mathcal{A}}(x,y,z)d\Omega 
=\int_0^{2\pi}\int_{-1}^1  \vec{\mathcal{A}}\left(x+n_x\Delta t, y+n_y\Delta t, z+n_z\Delta t\right)d\cos\theta d\phi \,,
\gal
where $\b n =( \sin\theta\cos\phi,\sin\theta\sin\phi,\cos\theta)$ is a unit vector along the radius of $S_{\b r}^{t-t_0}$ and we denoted $\Delta t=t-t_0$. 
Taking the time-derivative in \eq{a39} we obtain
\begin{subequations}
\bal
\vec A_2^\text{init}(\b r, t)&=\frac{1}{4\pi}\bigg\{\left[1+\frac{\sigma}{2}(t-t_0)\right]e^{-\frac{\sigma}{2}(t- t_0)}\oint_{S_{\b r}^{t-t_0}}  \vec{ \mathcal{A}}(\b r)d\Omega\label{a42}\\
&+(t-t_0)e^{-\frac{1}{2}\sigma(t- t_0)}\oint_{S_{\b r}^{t-t_0}} \vec{\mathcal{V}}(\b r)d\Omega\label{a43}\\
&+(t-t_0)e^{-\frac{1}{2}\sigma(t-t_0)}\oint_{S_{\b r}^{t-t_0}} (\b n\cdot\b \nabla ) \vec{\mathcal{A}}(\b r)d\Omega\bigg\}\,.\label{a44}
\gal
\end{subequations}
At this point it is instructive to confirm that Eqs.~\eq{a37},\eq{a42},\eq{a43} and \eq{a44} solve the initial value problem given by Eqs.~\eq{a7},\eq{a11} and \eq{a13}.  That \eq{a37} is a particular solution of \eq{a7} can be verified using \eq{a15} and \eq{a19}.  As $t\to t_0$ \eq{a42} satisfies the initial condition \eq{a11}, whereas \eq{a37},\eq{a43} and \eq{a44} vanish. Taking the time-derivative  and letting $t\to t_0$, we see that
\eq{a43} satisfies \eq{a13} whereas \eq{a37},\eq{a42} and \eq{a44} vanish. 

Integrals in \eq{a37} can be taken exactly. The retarded time $t'$ is determined from the condition $t-t'= R= |\b r-vt' \unit z|$, whose solution satisfying $t>t'$ reads 
\ball{d2} 
t'= \gamma^2\left( t-vz-\sqrt{(z-vt)^2+b^2/\gamma^2}\right)\equiv T\,.
\gal
The properties of the $\delta$-function allow us to write
\ball{d4}
\delta(t-t'-R)= \frac{\delta(t'-T) (t-T)}{\sqrt{(z-vt)^2+b^2/\gamma^2}}\,.
\gal
Substituting \eq{d4} and  $\vec j(\b r',t') = e\vec u \delta(\b b')\delta(z'-vt')$ into \eq{a37} yields
\ball{d6}
\vec A_2^\text{val}(\b r, t)=\frac{e\vec u}{4\pi}\frac{1}{\sqrt{(vt-z)^2+b^2/\gamma^2}}\exp\left\{-\frac{\sigma\gamma^2}{2} \left(-v(vt-z)+\sqrt{(vt-z)^2+b^2/\gamma^2}\right)\right\}\theta(T-t_0)\,.
\gal
The step function in the right-hand-side of \eq{d6} together with the retardation condition $t>t'=T$  guarantee that \eq{d6} vanishes as $t= t_0$.

The explicit form of the surface integral over the initial four-potential $\vec{ \mathcal{A}}(\b r)$ of the point charge appearing in \eq{a42} reads 
\bal
&\oint_{S_{\b r}^{\Delta t }} \vec{ \mathcal{A}}(\b r)d\Omega= \frac{e \vec u  }{4\pi} \oint_{S_{\b r}^{\Delta t}} \frac{\gamma d\Omega}{\sqrt{(x+n_x\Delta t )^2+(y+n_y\Delta t)^2+\gamma^2 (z-v t_0+n_z\Delta t)^2}} \,,\label{a45}
\gal
where $\vec u = (1,\b v)$, $\gamma=(1-v^2)^{-1/2}$. The other two surface integrals appearing in  \eq{a43},\eq{a44} can be cast in a similar form.  

Using the cylindrical coordinates $r,\phi, z$ and noting that for a point charge $A^b=0$, the non-vanishing components of the electromagnetic field are obtained as 
\ball{a47}
B_\phi = -\frac{\partial A^z}{\partial r},\,\quad  
E_r= -\frac{\partial{A^t}}{\partial r},\,\quad
E_z = -\frac{\partial{A^z}}{\partial t} - \frac{\partial{A^t}}{\partial z}\,.
\gal
 In the same vein as for the potential at $t>t_0$, it is convenient to separate the contributions of the valence current and the initial conditions, which will be referred to as the ``valence" and  the ``initial"  fields. Using the worldline coordinates $z_\pm = z\pm v(t-t_0)$ it is straightforward to see that the  longitudinal component $E_z$ is suppressed by a small factor $1/\gamma^2$ as compared to the transverse ones and can be ignored.

\medskip
The further computation of $\vec A_2$ is carried out numerically using Eqs.~\eq{d6} and \eq{a42}-\eq{a44} which are subsequently differentiated to obtain the fields \eq{a47}. Computation of the ``initial" part of the electromagnetic field turned out to be challenging  due to poor convergence of the surface integrals beginning at roughly $t\sim 5$ fm for  central collisions. Fortunately, the late-time behavior of this part of the electromagnetic field can be accurately described by the following analytical formula \cite{Tuchin:2015oka}
\ball{a60}
\vec A_2^{\text{init}}(\b r, t)\approx \frac{\kappa\gamma e \vec u}{4\pi}\int_0^\infty dk J_0(kb) e^{-k^2(t-t_0)/\sigma - k \gamma|z-vt_0|}\,,\quad t\gg t_0\,,
\gal
where $\kappa$ is a numerical coefficient of order unity fixed to provide smooth matching to the numerical results. Our procedure therefore is to compute the ``initial" contribution numerically through about the middle of the time-evolution, where it is superseded by a calculation based on the analytical formula \eq{a60}. The ``valence" part is computed numerically at all times. 

\begin{figure}
\begin{tabular}{cc}
 \includegraphics[height=5.5cm]{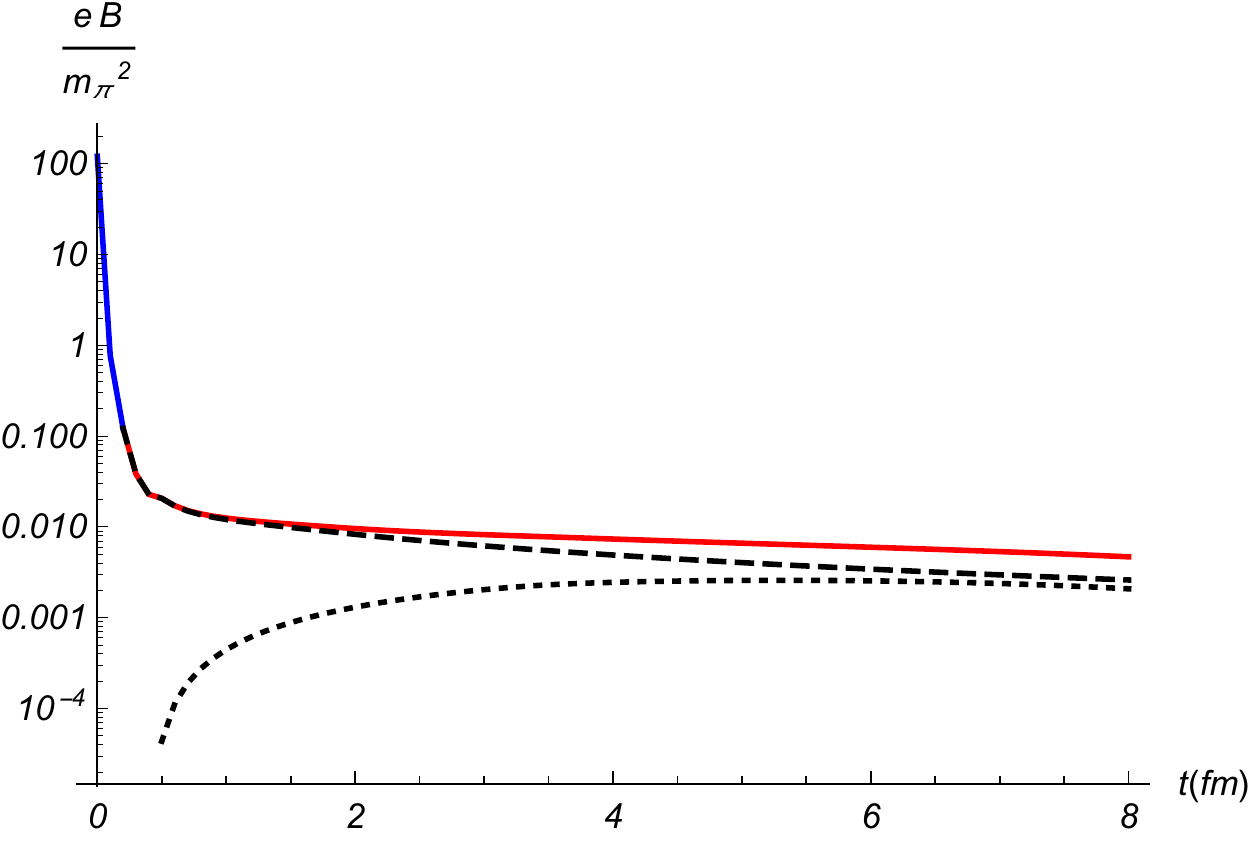} &
      \includegraphics[height=5.5cm]{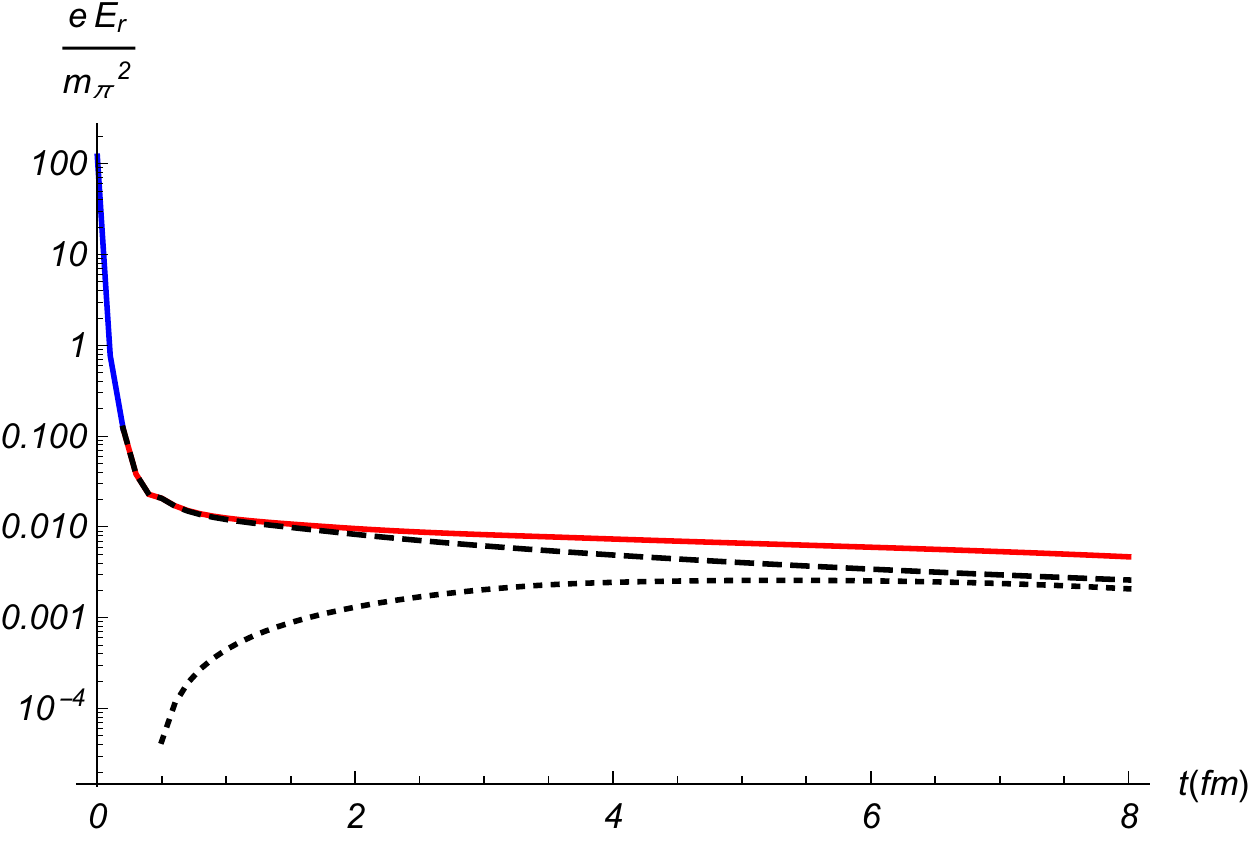}
      \end{tabular}
  \caption{(Color online) Time-evolution of the magnetic (left panel) and the radial electric (right panel) fields at $\sqrt{s_{NN}}=200$~GeV  at impact parameter $b=5$~fm. The observation point is $x=y=z=0$ (left panel) and   $x=z=0$, $y=1$ fm (right panel), with $xz$ being the collision plane. Blue line: $t<t_0$, red line: $t>t_0$, dashed line: the ``initial" field, dotted line: the ``valence" field.  }
\label{fig:B}
\end{figure}

\begin{figure}
\begin{tabular}{cc}
       \includegraphics[height=5.5cm]{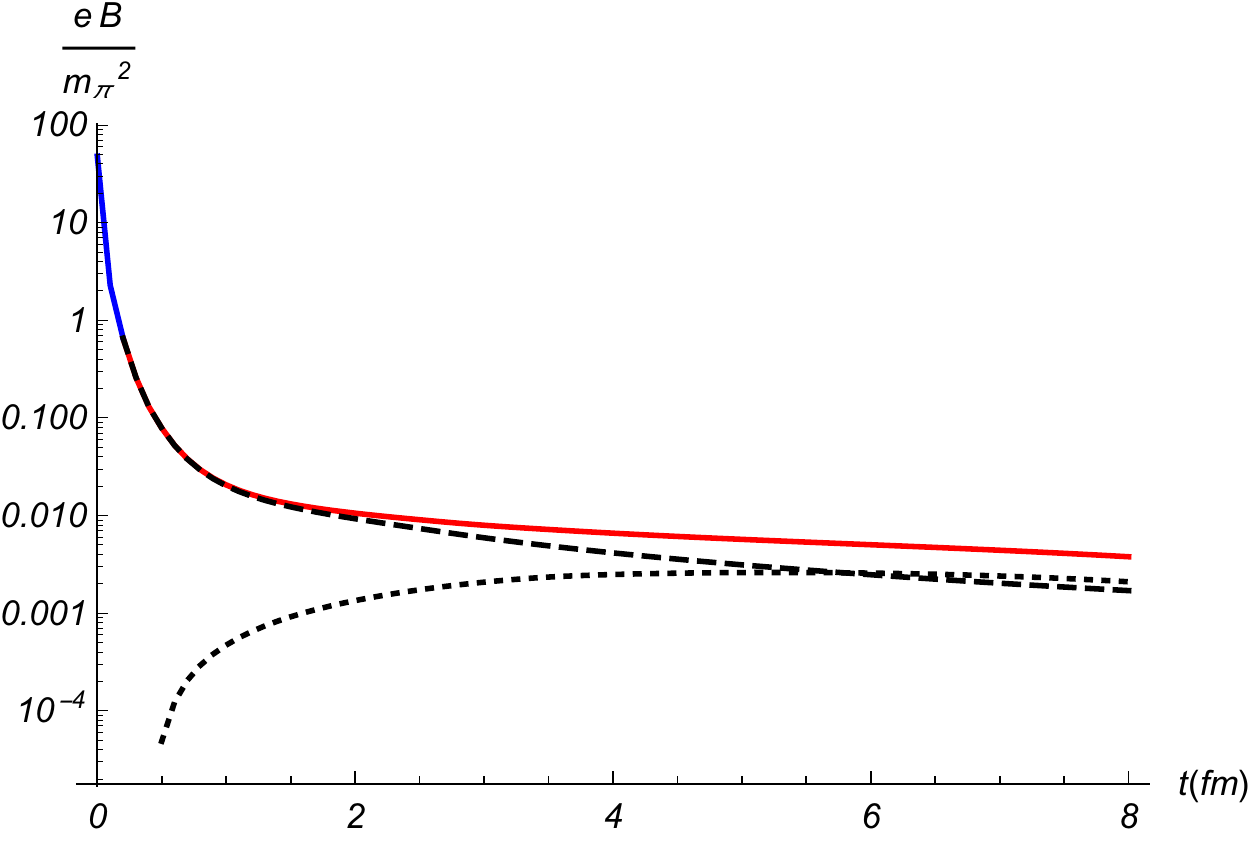} &
      \includegraphics[height=5.5cm]{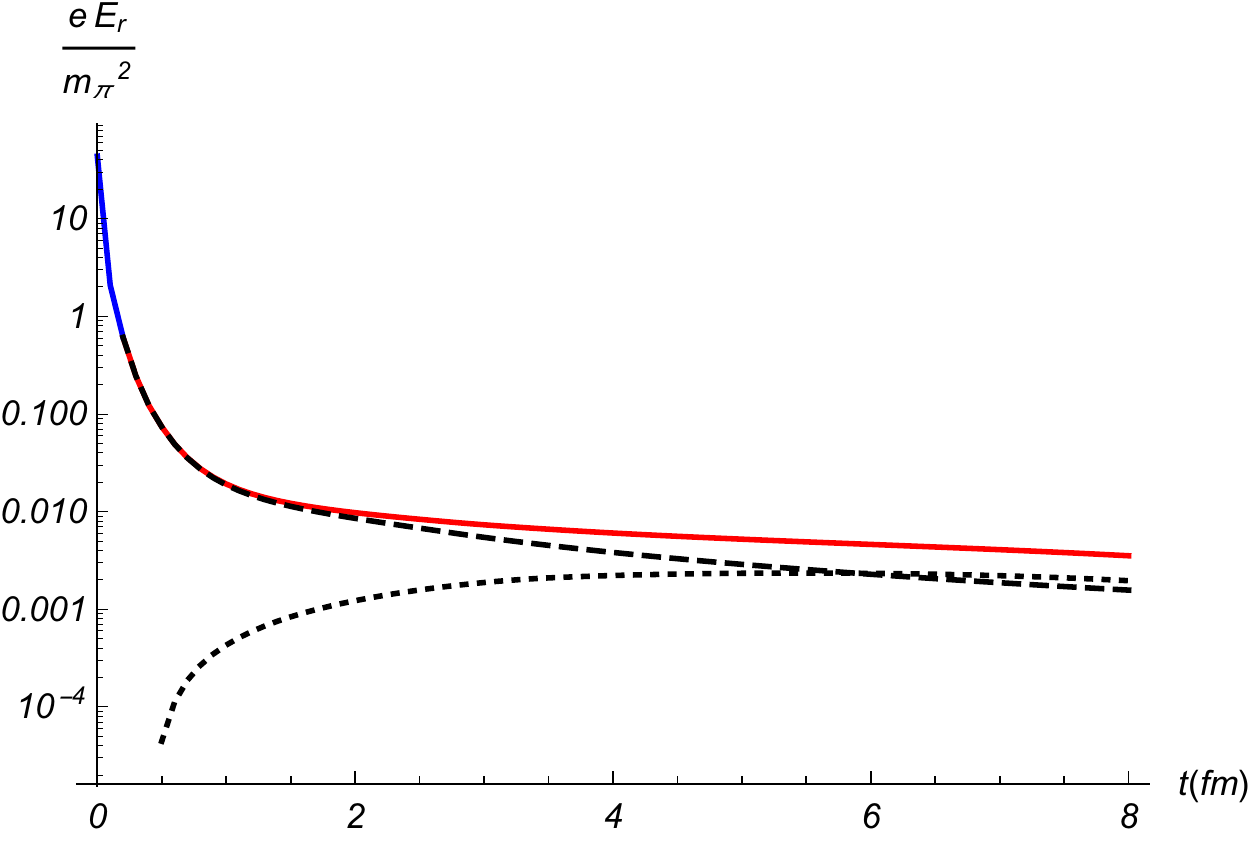}
            \end{tabular}
   \caption{(Color online) Time-evolution of the magnetic (left panel) and the radial electric (right panel) fields at $\sqrt{s_{NN}}=66.5$~GeV  at impact parameter $b=5$~fm. The observation point is $x=y=z=0$ (left panel) and   $x=z=0$, $y=1$ fm (right panel), with $xz$ being the collision plane. Blue line: $t<t_0$, red line: $t>t_0$, dashed line: the ``initial"  field, dotted line: the ``valence" field.}
\label{fig:B-g33}
\end{figure}

\begin{figure}
\begin{tabular}{cc}
       \includegraphics[width=7.5cm]{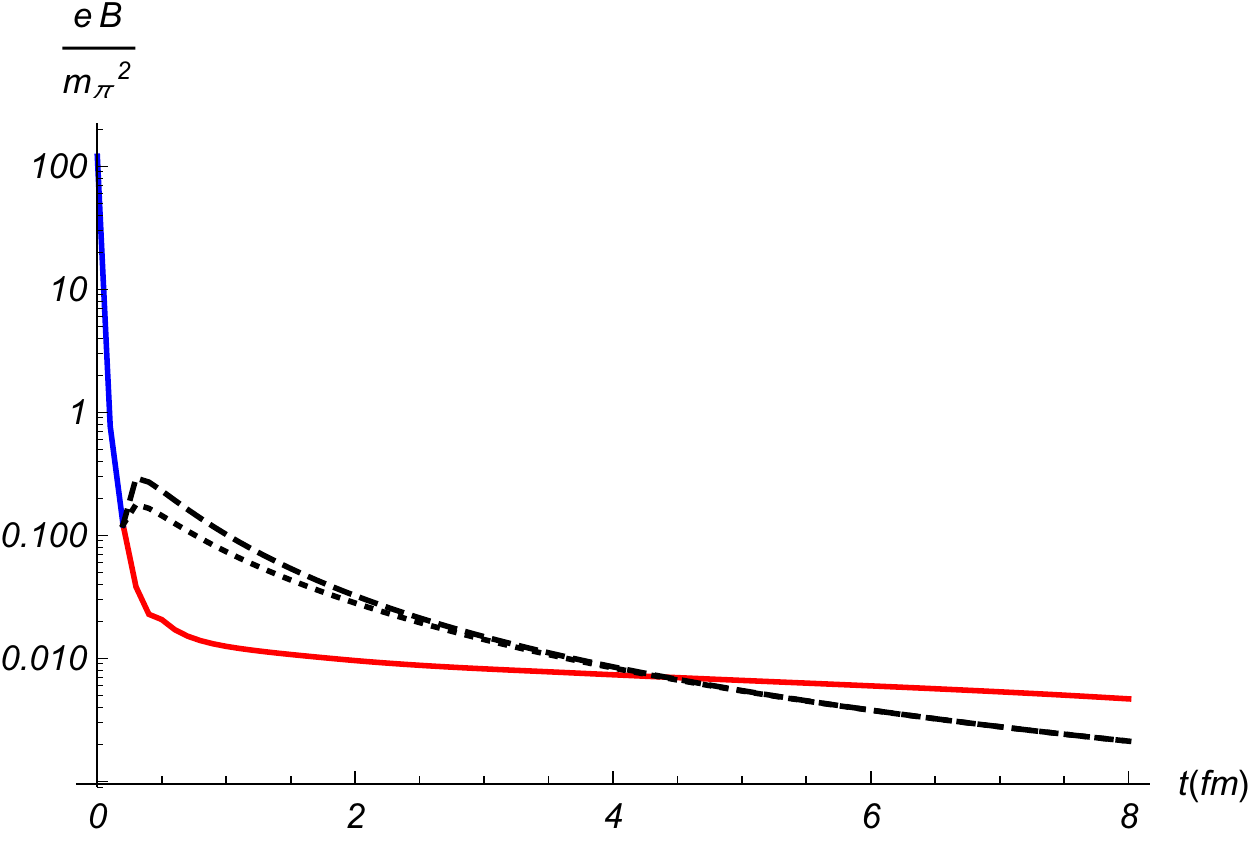} &
      \includegraphics[width=7.5cm]{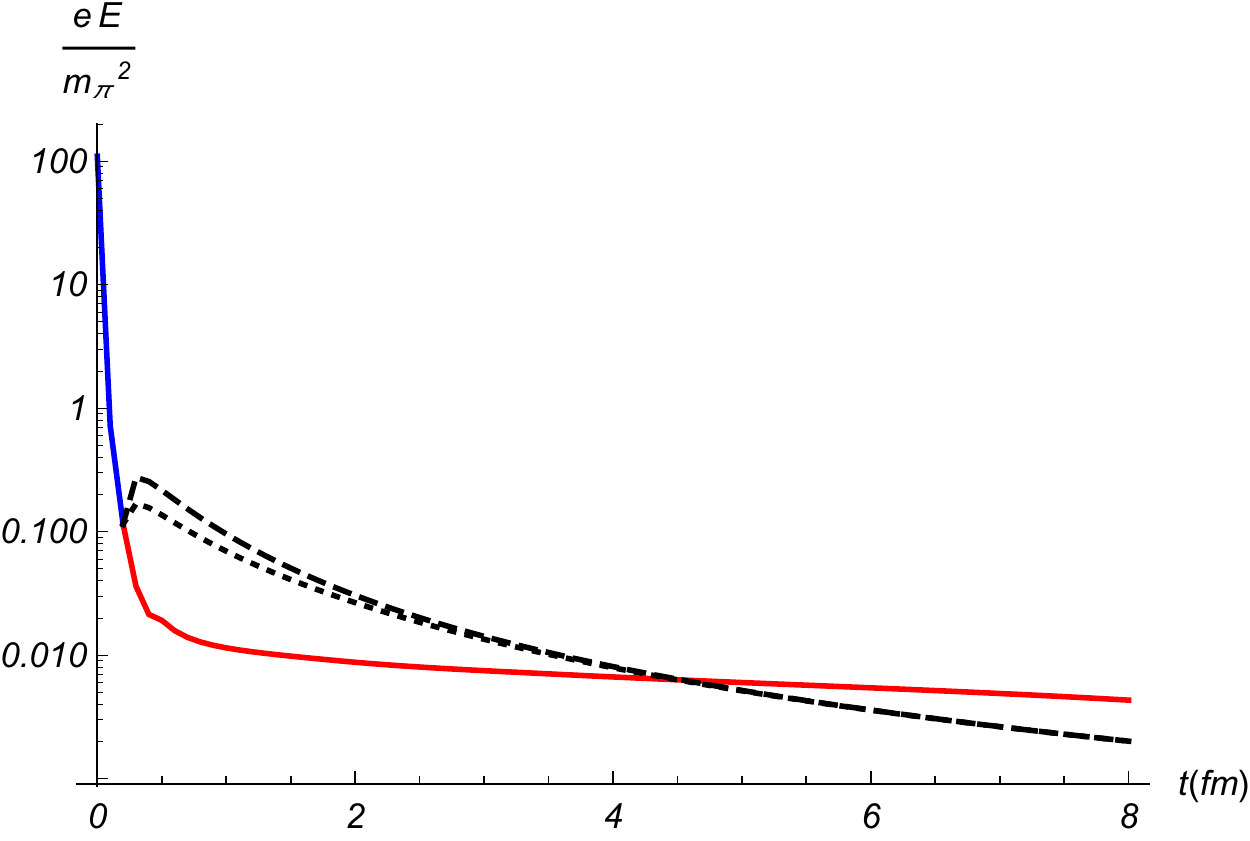}
      \end{tabular}
       \caption{(Color online) Time-evolution of the magnetic and the radial electric field  at $\sqrt{s_{NN}}=200$~GeV  at impact parameter $b=5$~fm in the phenomenologically favorable \emph{instantaneous} scenario (solid blue and red line) and the adiabatic scenario \eq{b5} with $\tau=1$~fm (dashed line) and $0.5$~fm (dotted line).  Observation point: $x=y=z=0$ in the left panel, $x=z=0$, $y=1$ fm on the central panel and right panels. }
 \label{fig:Er}
\end{figure}

\begin{figure}
\begin{tabular}{cc}
            \includegraphics[width=7.5cm]{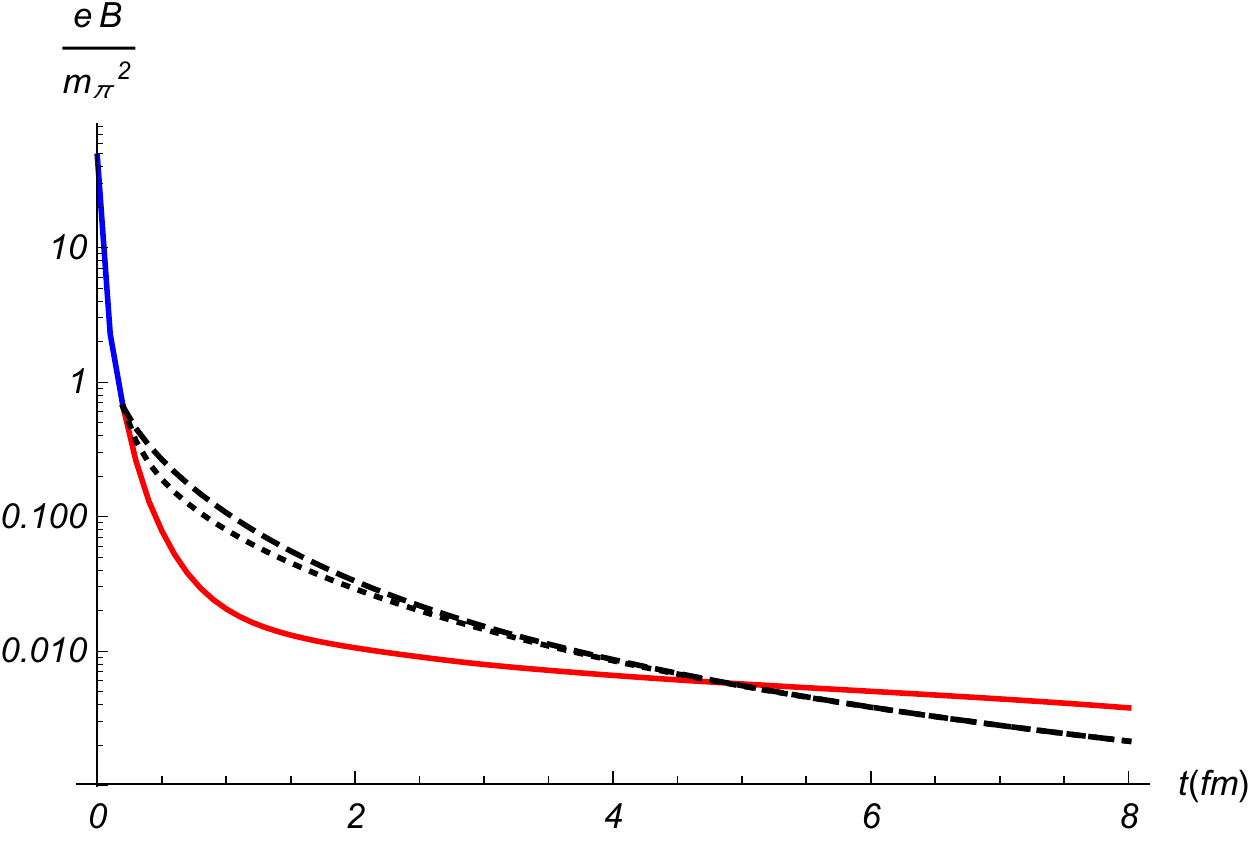} &
      \includegraphics[width=7.5cm]{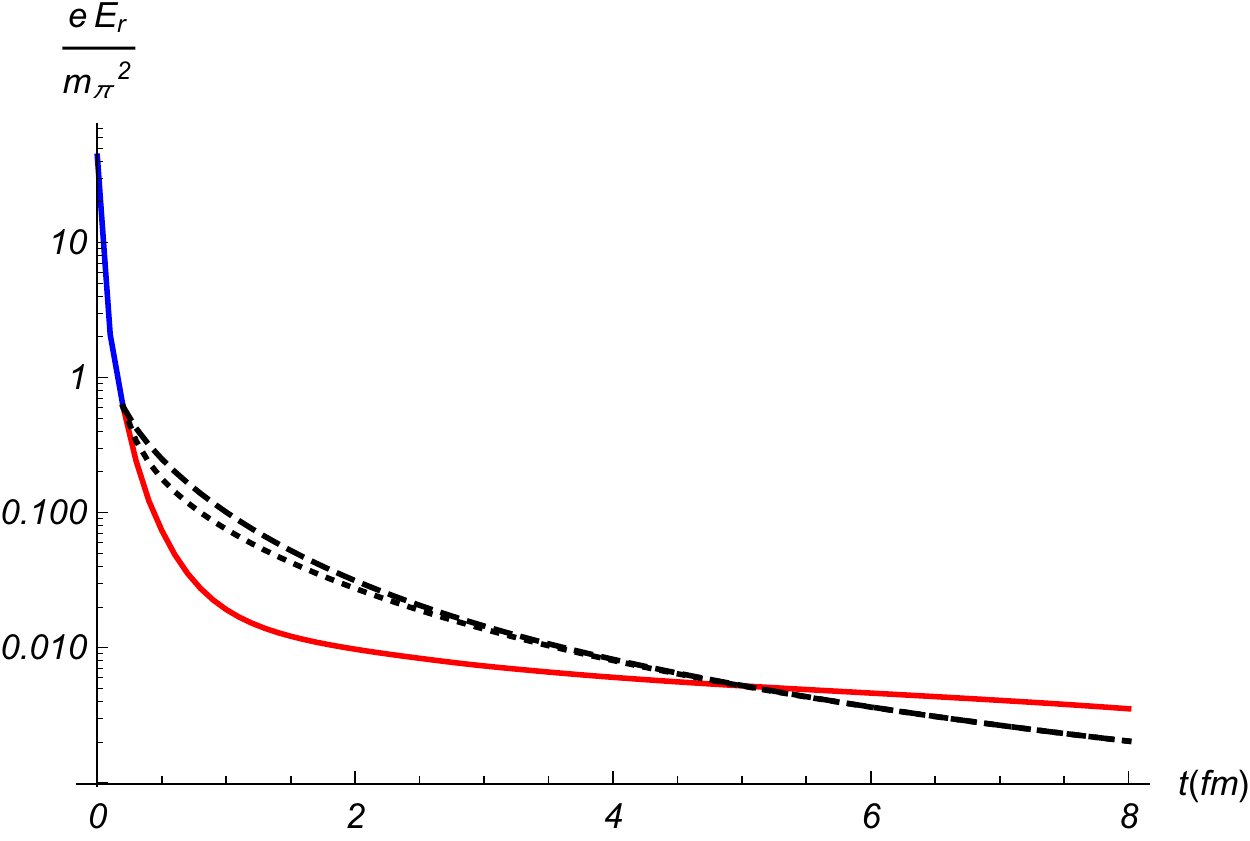}
      \end{tabular}
  \caption{(Color online). Time-evolution of the magnetic (left panel) and the radial electric (right panel) fields  at $\sqrt{s_{NN}}=66.5$~GeV  at impact parameter $b=5$~fm in the phenomenologically favorable \emph{instantaneous} scenario (solid blue and red line) and the adiabatic scenario \eq{b5} with $\tau=1$~fm (dashed line) and $0.5$~fm (dotted line).  Observation point: $x=y=z=0$ in the left panel, $x=z=0$, $y=1$ fm on the central panel and right panels.    }
\label{fig:Er-g33}
\end{figure}

The electromagnetic field of a heavy-ion is computed by summing up the contributions of the point-like charges. This is accomplished by integrating $\vec A_2 $ with the Woods-Saxon nuclear density. We assumed that the distribution of valence charges does not significantly change in the course of the collision. The total electromagnetic field after the heavy-ion collision is a sum of the electromagnetic fields created by each heavy-ion. It is exhibited in Figs.~\ref{fig:B},\ref{fig:Er}  and Figs.~\ref{fig:B-g33},\ref{fig:Er-g33}  for  Au-Au collisions at $\sqrt{s_{NN}}=200$ and 33 ~GeV respectively. The value of electrical conductivity of QGP is set at $\sigma=5.8$~MeV \cite{Aarts:2007wj,Ding:2010ga,Amato:2013oja}.

The most striking feature of these figures is very slow time dependence that sets in at about the same time as the conducting medium emerges. Such behavior was anticipated in \cite{Tuchin:2015oka} based on the following argument. The ``valence" part of the fields falls off at later times as $1/t^2$, whereas the ``initial" part as $1/t^{3/2}$. Hence the former dominates at earlier times while the later is more important at later times.  Their sum then tends to be a slower function of time than each of the two contributions in the intermediate region. Figs.~\ref{fig:B}, \ref{fig:B-g33} show that the ``initial" part quickly reaches the asymptotic behavior which is why we could use \eq{a60} to describe its evolution at later times. However, the ``valence" part displays monotonic growth indicating that it does not reach the asymptotic behavior at times relevant to the QGP phenomenology. This occurs because at $t=t_0$ the field created by the valence current in vacuum is assigned to the initial conditions so that the field of the valence current in QGP starts from zero. Due to the signal retardation, the field induced by the valence current in QGP builds up over roughly $2b$ starting from $t=t_0$\cite{Holliday:2016lbx,Peroutka:2017esw}. Since the latest evolution time we consider is 8~fm, our calculation never reaches the asymptotic late-time behavior.\footnote{We verified that at later times the asymptotic behavior is reached.} As a result the total electromagnetic field is nearly constant in QGP.  Although Figs.~\ref{fig:B}, \ref{fig:B-g33} display the field only at one point, its behavior at other points and at different impact parameters is qualitatively same.

\medskip

Our calculations thus far relied on an assumption---backed by the phenomenology---that QGP appears instantaneously as an electrically conducting medium at a certain $t_0$. We will refer to this model as the 
the \emph{instantaneous} scenario. The only parameter we used to describe the nuclear medium, its electrical conductivity $\sigma$, vanishes at $t<t_0$ and equals a constant at $t\ge t_0$.  To complete our discussion it is instructive to consider the opposite limit of an adiabatically emergent QGP in which case one can consider $\sigma$ as a slow function of time. In such an \emph{adiabatic} scenario the electromagnetic field can be obtained from \eq{d6} where $\sigma$ now is a function of time and $t_0\to -\infty$. This approach can be justified if the time interval over which the field significantly changes $\tau$ is  significantly larger than the QGP electric response time $\sigma/\gamma$. Time-variation of $\sigma$ can be captured by a simple two-parameter model 
\ball{b5}
&\sigma(t)= \sigma\left(1-e^{-(t-t_0)/\tau}\right)\theta(t-t_0)\,.
\gal
It takes into account the gradual emergence of the electrically conducting medium at early times, but ignores the effect of QGP at later times since it results only in a minor time-variation of the electrical conductivity. For example, in the Bjorken scenario \cite{Bjorken:1982qr} $\sigma(t)\sim t^{-1/3}$ \cite{Tuchin:2013ie}. The comparison of the two scenarios is presented in \fig{fig:Er} and \fig{fig:Er-g33}. It is seen that the electromagnetic field in the \emph{instantaneous} scenario is a much slower function of time than in the  \emph{adiabatic} one.

\medskip

In conclusion, we argue that the transverse electromagnetic field produced in Au-Au collisions at $\sqrt{s_{NN}}=200$~GeV is nearly time-independent after $t\sim 0.5$~fm/$c$, while  at $\sqrt{s_{NN}}=66.5$~GeV after $t\sim 1$~fm/$c$.

\acknowledgments
We thank Wyatt Peterson who assisted us with numerical calculations.  This work  was supported in part by the U.S. Department of Energy under Grant No.\ DE-FG02-87ER40371.


\end{document}